\documentclass[a4paper,11pt]{article}
\usepackage[top=3cm,right=3cm,bottom=2.5cm,left=2.5cm]{geometry}  
\usepackage[utf8]{inputenc}
\usepackage{setspace}
\usepackage{hyperref}
\usepackage{amsmath}
\usepackage{amsthm}
\usepackage{amssymb}
\usepackage{amsfonts}
\usepackage{dsfont}
\usepackage{mathrsfs}
\usepackage{graphicx} 
\usepackage{cite}
\usepackage{hyperref} 
\usepackage{cleveref}
\usepackage{tikz}
\usetikzlibrary{decorations.pathmorphing}
\newcommand{\ud}{\,\mathrm{d}}

\newcommand{\pd}{\partial}

\newcommand{\mb}[1]{\boldsymbol{#1}}

\DeclareMathAlphabet{\scr}{OT1}{rsfs}{m}{n}

\begin{document}
	\begin{center}
			\Large{\textbf{ Stress Tensor  on Null Boundaries }}\\
				\vspace*{1cm}
		{\large  \textbf{Ghadir Jafari}\footnote{ghjafari@ipm.ir}},\\ 
		{\normalsize \it  School of Particles and Accelerators, Institute for Research in Fundamental Sciences (IPM), P.O. Box 19395-5531, Tehran, Iran
		}
	\end{center}
	\vspace*{1cm}
	\spacing{1.5}
	
	\begin{abstract}
Using the Brown-York prescription for the definition of quasilocal gravitational energy-momentum tensor on a boundary and { also complete canonical structure on a null boundary which} has been found recently	 \cite{Aghapour:2018icu}, we propose a similar stress tensor on the null boundary. Then we exploit this stress tensor to compute the quasi-local energy and angular momentum  for some well-known gravitational solutions. We have found that in addition to reference spacetime method for regularizing total energy, in the case of null boundary we can add a possible counterterm so avoiding embedding difficulties.   
	\end{abstract}
\newpage
\tableofcontents	
	\section{Introduction}
	In general relativity (GR), defining a local energy for gravitational field is problematic (for a review on the subject see \cite{Szabados:2009eka}). The problem is rooted in the general covariance principle and the fact that the first derivative of metric can always vanish in a properly chosen local coordinate system. But in order to have a notion of local energy density with correct physical dimension, { such quantity could be } defined in terms of the metric and its first derivative. {Also, a} related notion to energy density in field theories is the action functional.  { In fact,} the standard covariant action in GR, {\it i.e.}, the Hilbert-Einstein (HE) action, contains second order derivatives of the metric, which is another consequence of the general covariance principle. As  { it} is well-known, this action does not have a well-posed variational principle and needs to be complemented with additional terms defined on the boundaries of the spacetime.  Brown and York\cite{Brown:1992br} { have pointed out that} having an action with a well-posed variational principle, by using Hamilton{-}Jacobi analysis, one can define the quasilocal energy of the system.

	A careful variation of the HE action in addition
	to suggesting the proper boundary term (boundary action),  to { ensure} a well-posed variational principle, also provide{s} dynamical degrees of the theory as well as its canonical structure (see {\it e.g.,} \cite{Padmanabhan:2014lwa}).  An important point, in finding the complete canonical
	structure by such a procedure, is the necessity { of the condition that one should} not suppress any degree of freedom beforehand by imposing
	restrictions.  The proper boundary action complementing the HE action, when the boundary is timelike or spacelike,  is the well-known Gibbons-Hawking-York (GHY) term \cite{Gibbons:1976ue,York:1972sj}.   
	 Applying variational principle in GR for null boundaries has been a subject of investigation in recent years\cite{Parattu:2015gga,Lehner:2016vdi,Hopfmuller:2016scf,Jubb:2016qzt,Hopfmuller:2018fni}.  In these papers, the authors {are} either interested in finding the proper boundary terms on null boundaries, and therefore ignored those terms that are fixed by boundary conditions, or imposed some restrictions on variations in such a way that the resulting canonical structure is not complete. One  of such restrictions is that  the variations keep the character of the boundary unchanged. But one of the metric degrees of freedom is responsible for variations which alter the boundary character. { This can be easily} seen as follows: If the boundary is specified by $ \phi=const $. for some scalar field $ \phi $, then the normal to the boundary is proportional to $ \pd_a \phi$. For  a null boundary, we have $ \pd_a\phi\pd^a\phi=0 $, while for a general metric variation we find:
	 \begin{equation}\label{var1}
	 \pd_a\phi\pd^a\phi\to\pd_a\phi\pd^a\phi-\delta g^{ab}\pd_a\phi\pd_b\phi\,.
	 \end{equation}
	Therefore, unless we set variations of metric in direction of $ \pd_a \phi$ to zero, the boundary doesn't remain null. 
	Recently we have shown that such variations appear in canonical structure\cite{Aghapour:2018icu}. 
{In addition, it has been } shown that, in order to preserve such variations, a general double-foliation framework is needed. 
	The conjugate momentum to these variations is a scalar $ \Xi $ (for definition see section \ref{NullST} or \cite{Aghapour:2018icu}).  In this article we will show that this scalar provides the quasilocal energy density on a null boundary 
%
	For {the} definition of quasilocal stress tensor on null boundary we follow the same method presented by Brown and York {in the case of} quasilocal stress tensor on timelike boundary which means { that the} derivative of {the} action with respect to {the} metric variations which are  tangential to the boundary\cite{Brown:1992br,Brown:2000dz}. 
	
The content of the paper is organized as follows. In the next section we first briefly review the original Brown-York construction of quasi-local gravitational energy-momentum tensor. In section \ref{NullST}{,} we follow the same analysis for the case where instead of timelike boundary we have a null one. Finally{,} in section \ref{charge} we calculate the energy and angular momentum in various space times using the proposed stress tensor. 

{Throughout this work}  we have set $ G=c=1 $. We do not use various indices for  specification of different parts of spacetime and use Latin indices $ \{a,b,c,\dots\} $ everywhere {; instead, different names are} given to objects when defined  on different  {structures}. For example {,} we use $ K_{ab} $ for extrinsic curvature of a spacelike hypersurface while $ \chi_{ab} $ is  {preserved} for extrinsic curvature of a timelike hypersurface.

\section{The Brown-York Tensor}  
  In this section we review the Brown-York \cite{Brown:1992br} definition of gravitational stress-tensor on the boundary. To illustrate the main idea beyond Brown-York definition of quasi-local energy momentum, it is useful  {to} start with an example in classical mechanics.
\subsection{The Hamilton-Jacobi  Method}
 Let $L(\mb{q}(t),\dot{\mb{q}}(t),t)$ be the classical Lagrangian for a particle. The action functional is  $I[\mb{q}(t)]=\int_{t_1}^{t_2}L(\mb{q}(t),\dot{\mb{q}}(t),t)\ dt$, for initial and final configuration  {, i.e.} $ (\mb{q}_1,t_1) $ and $ (\mb{q}_2,t_2) $, respectively.
  A general variation of this action for a given history is:
  \begin{equation}\label{varL}
  \delta I[\mathbf{q}(t)]=\int_{t_1}^{t_2}\big(\frac{\pd L}{\pd\mb{q}}-\frac{d}{dt}\frac{\pd L}{\pd\mb{\dot{q}}}\big)(\delta\mb{q}-\dot{\mb{q}}\delta t)dt+\frac{\pd L}{\pd\mb{\dot{q}}} \delta\mb{q}|_{t_1}^{t_2}-(\frac{\pd L}{\pd\mb{\dot{q}}}\mb{\dot{q}}-L)\delta t|_{t_1}^{t_2}
  \end{equation}
  Extremizing  the action  {and imposing} the boundary condition  {by} fixing $ \mb{q} $ and $ t $ at the end points,  {provides} the equations of motion that are given by the first term  in the integrand  { which are} known as the Euler-Lagrange equations. 
 {Moreover,} the  Hamilton{-}Jacobi principal function $ S(q_1,t_1;q_2,t_2) $ is defined as the value of the action for a solution $ \mb{q}(t) $ of the equation of motion from $ (\mb{q}_1,t_1) $ to $ (\mb{q}_2,t_2) $.
 
 {Thus,} according to \eqref{varL},  {the} derivative of  {the} principal function $ S $ with respect to $ \mb{q} $,  {i.e.} $ \frac{\delta S}{\delta\mb{q}} $, gives the canonical momenta $ \mb{p}=\frac{\pd L}{\pd \dot{\mb{q}}} $  {while the} derivative with respect to $ t $,  {i.e.} $ \frac{\delta S}{\delta t} $, yields  the minus of energy: $H=(\frac{\pd L}{\pd\mb{\dot{q}}}\mb{\dot{q}}-L)$.

An important { remark} is that in order  {to ensure that} the above procedure  works well, the variational  principle must be well-posed beforehand.  {Indeed,} a Lagrangian  with  {no} well-posed  variational principle does not lead to  { the}  correct relations for momenta and energy of the system. For example, the action $ S_1=\int \tfrac12 m \dot{x}^2 dt $ leads to the equations of motion and correct momentum and energy for a free particle.  { However, by considering another action as} $ S_2=S_1-\int \tfrac12 m \frac{d}{dt}(x\dot{x}) dt= -\int \tfrac12 m x\ddot{x} dt$ {, one could see that, although} the difference with  {the} first action is a total derivative and the equations of motion are unchanged,  {by varying this action one finds:}
\begin{equation}\label{dS2}
\delta S_2=-\int (m \ddot{x}) \delta x+[\tfrac12(\dot{x}\delta x-x \delta\dot{x})]_{t_1}^{t_2},
\end{equation}
which means that in order to get the equation of motion we need to determine both $ x $ and $ \dot{x} $ at both end points. This leads to inconsistency with second order equation of motion, because according to that we need to just fix the position at the ends.  {Therefore,} for this action   { the} variational principle is not well-posed\footnote{Of course this argument is valid if we are interested in  Dirichlet or Neumann boundary conditions. By choosing a Robin boundary conditions one could revive the variational principle.}.
As a consequence, the derivative of  {the} principal function does not lead to the correct definition for momentum and energy. However{,} we must note that  all total derivatives do not spoil the variational principle, {\it e.g.} changing the action by $ S\to S+\int_{t_1}^{t_2}  \tfrac{dh}{dt} dt$ for arbitrary function $ h(\mb{q}(t),t) $ is allowed. {Hence,} the action and {the} principal function are not unique{; as a result,} momenta and energy are also not exclusive.  {However,} this arbitrariness can be fixed by choosing  the zero point of energy, for example set the arbitrary function $ h $  so that it yields to zero energy for free particle at rest in the above example.

  \subsection{Stress tensor on timelike boundary}
{ Having} learned enough from the above simple mechanical  example, lets begin with HE action {in $d$ dimension}:
   	\begin{align}
  \mathcal S_{\scriptscriptstyle{EH}} =\tfrac{1}{16\pi} \int \ud^d x \sqrt{-g} \,R\,,
  \end{align}
{ in which $R$ is the Ricci scalar.} {By} varying the action, if we consider the boundary segments {to} be either {time-like or space-like, one gets}: 
  	\begin{align}
 \delta\mathcal S_{\scriptscriptstyle{EH}}&=\tfrac{1}{16\pi}\int_{\mathcal M}\!\!\! \ud^{d} x\sqrt{-g} G^{ab}\delta g_{ab}+\tfrac{1}{16\pi} \sum_{i}\big[2\delta\Big(\int_{\mathcal B_{i}}\!\!\! \ud^{d-1} x  \,\sqrt{|h|}\, K + \int_{\mathcal C_{i}}\!\!\! \ud^{d-2} x \,\sqrt{|q|}\vartheta\Big)\nonumber\\&+ \int_{\mathcal B_{i}}\!\!\! \ud^{d-1} x\sqrt{|h|}(K^{ab}-K\,h^{ab})\,\delta h_{ab}  + \int_{\mathcal C_{i}}\!\!\! \ud^{d-2} x \,\sqrt{|q|}\,\vartheta\,q^{ab}\,\delta q_{ab}\big],
  \end{align}
  where the sum is over each boundary segment $  \mathcal B_{i} $ and every corner $  \mathcal C_{i} $ at intersection of two neighboring  segments of boundary. {Moreover,} $ K $ is the extrinsic curvature of each segment and $ \vartheta $ is the angle or the boost parameter between segments {,}  depending on their character. {In addition,} $ \delta h_{ab}  $ is the variation of metric on each boundary {while} $ \delta q_{ab}  $ is the  variation on each co-dimension two joints.
  For  the details of the  calculations { and specially the method by which one can   take} into account the contribution of joints{,}  see Refs. \cite{Lehner:2016vdi,Jubb:2016qzt,Brown:2000dz} or \cite{Aghapour:2018icu}.  We see that {the} variation of HE action includes both the metric and its normal derivative (the extrinsic curvature) on the boundary, {thus, the} variational principle is not well-posed for this action if we demand the Dirichlet boundary conditions
  \footnote{For Neumann boundary condition in four dimensions, there is no need to any boundary term in the action to make  a well-defined variational principle, see for example \cite{Krishnan:2016mcj}. But usually in gravity one requires a  Dirichlet boundary condition for which the metric on the boundary is fixed. }. This variation also suggest{s} the correct action
   with well-posed variational principle {as}:
  \begin{equation}\label{well-posed}
\mathcal{S}=\tfrac{1}{16\pi} \int_{\mathcal{M}} \ud^d x \sqrt{-g} \,R-\tfrac{1}{8\pi}\sum_{i}\big[  \int_{\mathcal{B}_i} \ud^{d-1} x \sqrt{|h|}K+\int_{{\mathcal C}_i}\!\!\! \ud^{d-2} x \,\sqrt{|q|}\,\vartheta].
  \end{equation}
  
  \begin{figure}
  	\begin{center}
  		\includegraphics[scale=0.6]{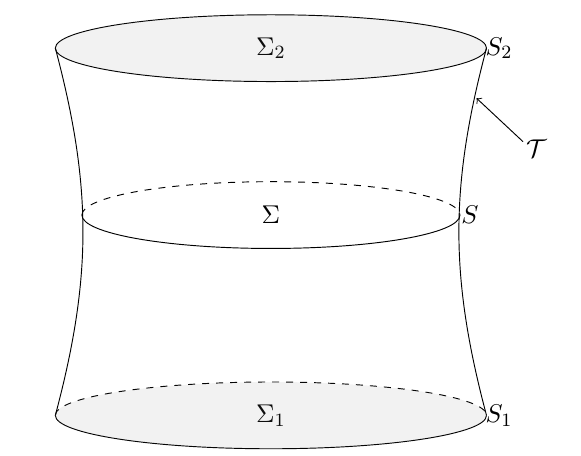}
  	\end{center}
  	\caption{Space time  region with timelike and spacelike boundaries}\label{fig1}
  \end{figure} 
  Now consider a region in {space-time}  as in fig \ref{fig1} which is generated by evolution of a spacelike surface $ \Sigma $ from
  $ \Sigma_1$ to $ \Sigma_2 $ and  {a} timelike boundary $ \mathcal{T} $. The intersection of $ \Sigma $ {leafs} with $ \mathcal{T} $ are co-dimension two surfaces $ S $ from $ S_1$ to $ S_2 $.
Now in this region of {space-time} we calculate the variation of the action \eqref{well-posed}, and impose the equations of motions {, i.e. $ G_{ab}=0 $. Then we may find:}
  	\begin{align}\label{deltaS}
  \delta\mathcal S&=\tfrac{1}{16\pi}\big[\int_{\Sigma_{1}}^{\Sigma_{2}}\!\!\! \ud^{d-1} xP^{ab}\,\delta h_{ab}  + \int_{\mathcal{T}}\!\!\! \ud^{d-1} x\Pi^{ab}\,\delta \gamma_{ab} +\int_{S_{1}}^{S_{2}}\!\!\! \ud^{d-2} x \,\sqrt{|q|}\,\vartheta\,q^{ab}\,\delta q_{ab}\big].
  \end{align}
{Here,}  the symbol $ \int_{\Sigma_{1}}^{\Sigma_{2}} $ is a  shorthand for $ \int_{\Sigma_2}-\int_{\Sigma_1}$, $ h_{ab} $ is the metric on each spacelike boundary and $ \gamma_{ab} $ is the induced metric on the timelike boundary $ \mathcal{T} $. $ P^{ab} $ and $ \Pi^{ab} $ are  respectively the gravitational momenta of $ \Sigma $ and $ \mathcal{T} $:
  \begin{align}\label{momentum}
  &P^{ab}=\sqrt{h}(K_{ab}-K h_{ab})|_{\Sigma},\nonumber\\
  &\Pi^{ab}=\sqrt{-\gamma}(\chi_{ab}-\chi \ \gamma_{ab})|_{\mathcal{T}},
  \end{align}
  where $ K_{ab} $ and $ \chi_{ab} $ are their corresponding extrinsic curvatures. 
 In the original work of Brown and York\cite{Brown:1992br}, they assumed the boundaries to be orthogonal, {thus,} the contribution of the joints was missing. {However,} after the work of Hayward\cite{Hayward:1993my}, { who emphasized}  the importance of this term for the variational principle to be well-posed, this term appeared in later works \cite{Hawking:1996ww,Booth:1998eh,Brown:2000dz}.
 
\begin{figure}
	\begin{center}
		\includegraphics[scale=0.35]{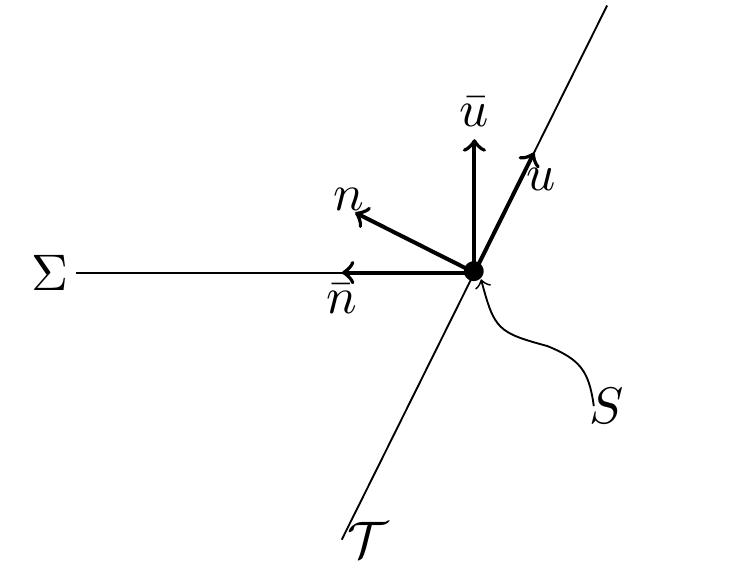}
	\end{center}
	\caption{Non-orthogonal boundaries.}\label{fig11}
\end{figure}

If we ignore these joint{s} and consider the boundaries to be orthogonal, then in {an} analogous way to the mechanical example one may define the gravitational canonical momentum as the derivative of {the}  principal functions with respect to the induced metric on spacelike segments of {the} boundary and gravitational energy-momentum-stress tensor as {the} derivative   with respect to the induced metric on timelike segment. In fact{,} the  $ \Pi^{ab} $  has the same expression as for  ADM canonical momentum. {Thus,} the energy-momentum-stress  tensor (or just the stress tensor for abbreviation) will be defined as:
  \begin{equation}\label{stress}
  T^{ab}=\frac{2}{\sqrt{-\gamma}}\frac{\delta S}{\delta \gamma_{ab}}=\tfrac{1}{8\pi}(\chi_{ab}-\chi \ \gamma_{ab}).
  \end{equation}
  To define conserved quantities{,} now we choose a spacelike co-dimension two surface $ S $ in $ \mathcal{T} $ with unit timelike normal $ u^{a} $.  Then{,} the metric $ \gamma_{ab} $ is further decomposed as $ \gamma^{ab}=q^{ab}-u^au^b $. Here{,} $ u^{a} $ defines local {the} flow of time in  $ \mathcal{T} $. { Moreover, according  to Brown-York \cite{Brown:1992br}}
  for an isometry in {the} boundary  generated by the Killing vector $ \xi^a $, the conserved charge associated to this symmetry  is defined by:
  \begin{equation}\label{Qcharge}
  Q_{\xi}=\int_{S}d^{d-2}x\sqrt{q}\ T_{ab}\,u^a\xi^b.
  \end{equation}
 In fact{,} there is some points regarding to the expression \eqref{stress} and \eqref{Qcharge}. The first one is that, as we pointed out {before}, these expressions are not unique.  One can append a subtraction term $ S_0 $  to the boundary action without affecting variational problem when $ S_0 $ depends on fixed boundary data, $ S_0=S_0(h_{ab},\gamma_{ab}) $, which leads to ambiguities in  {the definitions} of energy and momenta.{ According to Brown and York interpretation these ambiguities are consequence of freedom to choose the zero point of energy and redefine system momenta with canonical transformation \cite{Brown:1992br,Brown:2000dz}}. On the other hand, Eq. \eqref{Qcharge} leads to infinities when calculated for large spheres in  general systems in spacetime. The Brown-York { proposal is}  to choose subtraction term $ S_0 $  such that the modified action $ S-S_0 $  leads to zero energy for flat spacetime. {Therefore,} the zero point is chosen to be  { the } flat spacetime  {while} the scheme for other geometries is to embed  {their} boundary in flat spacetime. {Thus,} the modified expression for the stress tensor will be
  \begin{equation}\label{stress1}
 T^{ab}=\frac{2}{\sqrt{-\gamma}}\frac{\delta S}{\delta \gamma_{ab}}-\frac{2}{\sqrt{-\gamma}}\frac{\delta S_0}{\delta \gamma_{ab}}=\tfrac{1}{8\pi}(\chi_{ab}-\chi \ \gamma_{ab})-\frac{2}{\sqrt{-\gamma}}\frac{\delta S_0}{\delta \gamma_{ab}}.
 \end{equation}
  As another remark{,} if we want to consider the effect of joints or non-orthogonal boundaries, we must care about the components of $ \delta \gamma_{ab} $ contained in $ \delta q_{ab} $ appeared in the last term of  \eqref{deltaS}{. Also,} we must be careful  { about the question that with respect to which observer we desire to} calculate the quasi-local quantities. Note that when the boundaries are non-orthogonal, the Eulerian observers orthogonal to $ \Sigma $ constant are different from  the observers orthogonal to $ S $ constant in the boundary $ \mathcal{T} $ . In these cases, a further decomposition of  the induced metric $\gamma_{ab} $ with the assistance of the vector $ u_a=N\nabla_at $ is required, where $ t $ is a foliation of $ \mathcal{T} $. By this decomposition  we get \cite{Brown:2000dz}:
   \[ \delta\gamma_{ab}=\delta q_{ab}-\frac{2}{N} u_{(a}  \delta\,V_{b)} -u_a u_b \ \frac{\delta N}{N} , \]
where $ N $ and $ V^a $ are lapse and shifts of decomposition. The components of 
  $\frac{ \delta S}{\delta\gamma_{ab}} $ can be calculated as 
 \begin{align}
\epsilon&\equiv{}u^{a}u^{b}T_{ab}=-\frac{1}{\sqrt{q}}\frac{\delta{\mathcal{S}}}{\delta{N}}\label{energydensityt},\\
j^a&\equiv q^{ac}u^{b}T_{cb}=-\frac{1}{\sqrt{q}}\frac{\delta{\mathcal{S}}}{\delta{V_{a}}}\label{momentomdesityt},\\
s^{ab}&\equiv q^{ac}q^{bd}T_{cd}=\frac{2}{N\sqrt{q}}\frac{\delta{\mathcal{S}}}{\delta{q}_{ab}}\label{Sstrest},
\end{align} 
{ which are known respectively as} the  quasilocal energy density, tangential momentum density and spatial stress.
The  details of calculations for different observers can be found in \cite{Brown:2000dz,Booth:1998eh}. In these cases it has  been shown that a double foliation of space-time is a natural setup for calculations as described in appendices of \cite{Brown:2000dz}. 
There, the relation between quantities, as measured by different observers, has been obtained.
The relation{s} between two sets of  normals   to $ S $ as depicted in Fig.\ref{fig11} is:
\begin{align}\label{boost}
&\bar{n}_a=\gamma( n_a- v u_a)\\
&\bar{u}_a=\gamma( u_a- v n_a)
\end{align}

The quasilocal energy density   associated with the two surfaces $ S $ as seen by the  observers orthogonal to $ \Sigma $ constant is\cite{Brown:2000dz}:
\begin{equation}\label{energy}
\epsilon=\frac{1}{8\pi}(\gamma \mathbf{k}+\gamma v \mathbf{l})-\frac{2}{\sqrt{-\gamma}}\frac{\delta S_0}{\delta \gamma_{ab}}u^au^b,
\end{equation}
where $ \mathbf{k}_{ab}=-q_a^cq_b^d\nabla_c\bar{n}_d  $ and $ \mathbf{l}_{ab}=-q_a^cq_b^d\nabla_c\bar{u}_d  $ are defined  so that $\mathbf{k}$ and $\mathbf{l}$ are their trace, respectively.
In the case for which two observers are at rest with respect to each other {i.e. $ v=0 $ or, in other words, the boundaries are orthogonal,} the quasilocal energy density becomes:
\begin{equation}\label{QLE}
\epsilon=\frac{1}{8\pi}(\mathbf{k}-\mathbf{k}_0),
\end{equation}
where $ \mathbf{k}_0 $ is the corresponding  extrinsic curvature as embedded in flat space. So, the total quasilocal energy on $ S $ becomes\cite{Brown:1992br,Brown:2000dz}:
\begin{equation}\label{TQLE}
E=\frac{1}{8\pi}\int_{S}\ud^{d-2} x\sqrt{q}(\mathbf{k}-\mathbf{k}_0).
\end{equation}
Evaluating the above integral {provides} an expression for $ E $ as a function of $ r $, $ E=E(r) $. {However,} we must note that{  because of the Eq. \eqref{energy}}, in general the form of this function varies for different observers. For example, in the case of   Schwarzschild black hole the function for static, radially infalling  and boosted observers has been obtained in \cite{Booth:1998eh}. 

{As the last point,} let us note that for the above relations {, and for different observers,} it is always assumed that these observers do not move in null geodesics. In fact, null observers can not be contained in the above  setup because the normal vectors (or four-velocities) are always normalized to unity.
 As the name indicate{s}, { due to the fact that the  norm vanishes  for null observers,} normalization does not make sense  { in this case.}  { In fact, here,} the normal vectors or four-velocities are both normal and tangent to the hypersurface, the induced metric becomes degenerate, and the usual extrinsic curvature does not make sense.

        In the next section we will use standard treatment for null hyper-surfaces. The resulting stress tensor can {provide} the quasi-local quantities as measured by null observers. 
       \subsection{Asymptotic AdS spacetime and counterterm method} 
   Finding proper subtraction term by embedding in reference spacetime is a difficult task. In fact{,} it is not possible to embed a boundary with an arbitrary 
       intrinsic metric in the reference spacetime. In the case of asymptotically  AdS spacetimes{,} there is an attractive proposal without necessity of embedding in  reference spacetime as proposed  in \cite{Balasubramanian:1999re}. This approach is inspired from AdS/CFT duality \cite{Maldacena:1997re,Witten:1998qj} by interpreting  divergences which appear in stress tensor when the boundary is moved to infinity as dual to standard ultraviolet divergences of quantum field theory {; then,} they can be removed by adding  local counterterms to the boundary action. For instance for $ \mathrm{AdS}_4 $ the counterterm Lagrangian in the boundary proposed to be:
       \begin{equation}\label{Ctrem}
       L_{ct}=-\frac2{\ell}\sqrt{-\gamma}(1-\frac{\ell^2}{4}\mathcal{R}),
       \end{equation}
       where $ \mathcal{R} $ is the scalar curvature of induced metric on the boundary, and $ \ell $ is the AdS radius. Adding this term to the usual GHY term in the timelike boundary, then variation of action yields to a regularized stress tensor  such that energy becomes finite at the  $ r\to\infty $ limit\footnote{In \cite{Miskovic:2014zja} a topological method is proposed for finding  standard
       	counterterm series of AdS gravity.}. The main {advantage} of this method, beside the fact that counterterms are covariant and do not spoil the variational principle, is {the} lack of embedding difficulties in BY method. {However,} these counterterms are known just for asymptotically AdS spaces. For asymptotic flat space, { since there is no}  length scale $ \ell $, finding such counterterms is problematic. { Also, the limit $ \ell\to\infty $ in Eq.} 
       \eqref{Ctrem} does not lead to a unique covariant expression \cite{Costa:2013vza}.
       
  \section{Hamilton-Jacobi analysis on null boundary}\label{NullST}
\begin{figure}
 		\begin{center}
 			\includegraphics[scale=0.6]{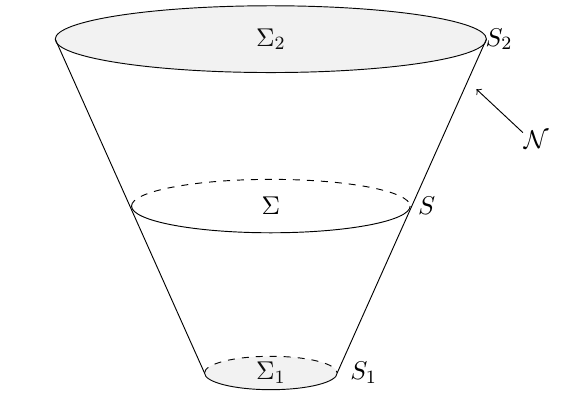}
 	\end{center}
 	\caption{Space-Time  region with a null and space-like boundaries}\label{fig2}
 \end{figure}

 In this section we are going to repeat the calculation of previous section when instead of a timelike boundary we have a null one. The corresponding spacetime region is illustrated in Fig.\ref{fig2}. In order to calculate {the} variation of the action in this region we need to express surface divergences in { the} variation of HE action in terms of geometric objects of spacelike surfaces{, i.e. $ \Sigma_1 $, $ \Sigma_2 $ and the null boundary $ \mathcal{N} $}. Variation on $ \Sigma_1 $ and $ \Sigma_2 $ is similar to the previous section{; however,} for a null surface (because of degeneracy of induced metric and divergence of extrinsic curvature) the calculation is completely different. In this section{,} we first introduce the basic tools for general investigation of null boundaries (without any gauge fixing) by introducing a general double foliation. Then{,} we find canonical momenta in this hypersurface { so that by varying the} action with respect to the metric components {on} this {hyper}surface we find stress tensor on {the} boundary. 
 
 \subsection{The set up}
 A segment $\mathcal N$ of the spacetime boundary, characterized by $\phi_0=0 $, is  called a null hypersurface if $\nabla_a\phi_0 \nabla^a\phi_0 = 0$. This feature of null boundary  indicates that the normal vector to the null surface is also tangent to it. This property is the origin of some difficulties when dealing with such hypersurfaces;  because, as a consequence, the induced metric becomes degenerate and therefore constructing a projector to the null surface just from its normal is not possible. One standard remedy to this problem is to introduce an auxiliary null vector $k^a$ which lays out of the hypersurface and therefore $\ell_a \, k^a \ne 0$, when $\ell_a$ is the null normal to the boundary, e.g., $ \ell_a\propto\nabla_a\phi_0 $ on the boundary. {For} more details about the geometry of null hypersurfaces we refer the interested reader to \cite{Gourgoulhon:2005ng,Torre:1985rw}. 
 
 {By defining $\ell_a$  as the normal to the null boundary,  we} introduce the auxiliary null form $k_a$ and take the normalization of these null forms to be everywhere {as}:  
 \begin{align}
 \ell_a\,\ell^a =0\ ,\ k_a\,k^a = 0 \quad\text{and}\quad \ell_a\,k^a=-1\,. \label{lknormalization}
 \end{align} 
 With the aid of $\ell_a$ and $k_a$, we can define the projector as:
 \begin{align}
 q^a{}_b = \delta^a{}_b + \ell^a\,k_b + k^a\,\ell_b
 \end{align}
 This projector is not in fact a projector on null surface{; instead,} it essentially projects spacetime vectors onto the co-dimension two surface $ { S}$, to which $\ell_a$ and $k_a$ are orthogonal.

 A systematic approach to define these co-dimension two surfaces is to use a double-foliation by two scalar fields $ (\phi_0,\phi_1) $. The intersection of level surfaces of $\phi_0$ and $\phi_1$ are the co-dimension two surfaces $ S$. In this foliation $ \ell_a $ and $ k_a $ can be expanded generally as:  
 \begin{align}
 &\ell_a=A \,\nabla_a\phi_0+B\, \nabla_a\phi_1 \label{lfoliation}\\
 &k_a=C\, \nabla_a\phi_0+D\, \nabla_a\phi_1 \label{kfoliation}
 \end{align}
 Three {out} of four coefficients in the above expansions are determined by normalization conditions \eqref{lknormalization} and one remains free, due to the re-scaling or boost gauge freedom $(\ell_a \to \alpha\ell_a,k_a\to\frac1\alpha\kappa_a)$. On the boundary, we set the coefficient $ B=0 $, so that the boundary is a level surface of $\phi_0$ and we have $\ell_a  \overset{\mathcal B}=A\,\nabla_a\phi_0  $, and it is null because $ \nabla_a\phi_0\nabla^a\phi_0\propto \ell^a \ell_a=0 $. Note that $ \ell_a $ and $ k_a $ are form fields defined in whole {space-time} and everywhere we have $ \ell^2=k^2=0 $, whereas  $ \nabla_a\phi_0\nabla^a\phi_0=0 $ is satisfied just on the boundary.  In fact, in general we have $ \nabla_a\phi_0\nabla^a\phi_0=\frac{2B D}{AD-BC}  $ from  which one could specify the location of null boundary as $ B=0 $.  The other point regarding to this foliation comparing to single foliation is that the vectors $ \ell_a $ and $ k_a $ are not in general hypersurface orthogonal. In fact using \cref{lfoliation,kfoliation}, one can easily evaluate $  \ell_{[a}\nabla_b\ell_{c]}  $ or $  k_{[a}\nabla_bk_{c]}  $, which for general value of functions $ \{A,B,C,D\} $ do not vanish. So according to Frobenius theorem, vectors $ \ell_a $ and $ k_a $ are not in general hypersurface orthogonal.
  
 In this double foliation framework,  the spacetime metric becomes:
 \begin{align}\label{metric}
 g_{ab}\,dx^a\,dx^b = H_{ij}\,d\phi^i\,d\phi^j + q_{AB}(d\sigma^A + \beta_i^A\,d\phi^i)(d\sigma^B + \beta_j^B\,d\phi^j)\,,
 \end{align}
 { in which} $\{i,j\}\in\{0,1\}$ whereas $\{A,B\}\in\{2,\ldots ,d-1\}$. Here $\sigma^A$ are coordinates on co-dimension two surface $ S$ and $\beta^A_i$ are shift vectors. The normal metric $H_{ij}$ consists of lapse functions as
 \begin{align}
 H_{ij} = -\left(
 \begin{array}{cc}
 2  A {C} & {B} {C}+{A} {D} \\
 {B} {C}+{A} {D} & 2 {B} {D} \\
 \end{array}
 \right).
 \end{align}
 By covariant differentiation of vectors $ \ell_a $ and $ k_a $ and projecting them in different directions, using $ q^a_b $, $ \ell^a $ and $ k^a $, we can  define the following geometric objects from $ \nabla_a\ell_b $ and $ \nabla_a\ell_b $ :
 \begin{align}
 \nabla_a \ell_b = -\Theta_{ab} - \omega_a\,\ell_b - \ell_a\,\eta_{b} - k_a\,a_b + \kappa\,k_a\,\ell_b - \bar{\kappa}\,\ell_a\,\ell_b \label{delldecomposition}\\
 \nabla_b k_b = -\Xi_{ab} + \omega_a\,k_b - k_a\,\bar{\eta}_b - \ell_a\,\bar a_b - \kappa\,k_a\,k_b + \bar{\kappa}\,\ell_a\,k_b \label{delkdecomposition}
 \end{align}
 These relations are generalizations of the relation $ \nabla_an_b=-K_{ab}+n_a a_b $ to current case  where  decomposition has been done with two null vectors. The definitions are as follows:
 \begin{equation}\label{BGOdef}
 \begin{aligned}
 &\Theta_{ab} =- q^c{}_a\,q^d{}_b\,\nabla_a\ell_b \ &&, \quad \Xi_{ab} = -q^c{}_a\,q^d{}_b\,\nabla_ak_b, \\
 &\eta_a =  q^c{}_a\,k^b\,\nabla_b\ell_c \ &&, \quad \bar{\eta}_b =  q^c{}_a\,\ell^b\,\nabla_b k_c, \\
 &\omega_a =  q^c{}_a\,k^b\,\nabla_c\ell_b = -q^c{}_a\,\ell^b\,\nabla_c k_b, \\
 &a_a = q^c{}_a\,\ell^b\,\nabla_b\ell_c \ &&, \quad \bar a_a = q^c{}_a\,k^b\,\nabla_b k_c, \\
 &\kappa = \ell^a\,k^b\,\nabla_a\ell_b =-  \ell^a\,\ell^b\,\nabla_a k_b \ &&, \quad \bar\kappa = k^a\,\ell^b\,\nabla_a k_b = -k^a\,k^b\,\nabla_a\ell_b,
 \end{aligned}
 \end{equation}
 where $\Theta_{ab}$ and $\Xi_{ab}$ are extrinsic curvatures of $S$ { while $\omega_a$, $\eta_a$ and $\bar{\eta}_a$ are twists.} { In addition, $a_a$ and $\bar a_a$ are tangent accelerations of $\ell^a$ and $k^a$ to $ S$, respectively. Moreover,} $\kappa$ and $\bar\kappa$ are in-affinity parameters\footnote{The quantity $\kappa$ is called surface gravity in the case of a black hole horizon null surface.}.
 
 As a side remark{,} let us point out that this general double foliation  described above is different from double null  foliation 
 as considered by various authors \cite{Brady:1995na,dInverno:2006wzl} . {Although, the double null  foliation is} useful for initial value problem  as shown by Sachs \cite{Sachs:1962zzb}, { one could show that for variational principle it suffers from having  a partial  gauge fixing condition which is a disadvantage\cite{Goldberg:1992st}.} This is because  requiring
 that the level sets of a coordinate $ \phi $ being null, fixes one of the metric components:
 \[ g^{\phi\phi}=g^{-1}(d\phi,d\phi)=0. \]
 The double null foliation is {a special case of the} above set up if we set $ B=C=0 $ in whole spacetime. 
 If fact{,} we will see in the next section that {the} variation of such components is important for finding complete canonical structure, and more importantly the canonical momenta of such variations is the quasilocal energy density of system.  
 
 \subsection{Variation  of Hilbert-Einstein action}
 %
 The main point in { calculating the variations} on a null surface is that the operators $ \delta $ for variations and the covariant derivative $ \nabla $  are defined in spacetime, whereas the condition for boundary, in the above setup $ B=0 $, is valid only on the null boundary. {Thus,} we must first apply these operators and then impose the condition $ B=0 $. For example consider variation of the vector $ \ell_a $  given by Eq. \eqref{lfoliation}. By variation of this vector on the boundary, one finds $ \delta\ell_a=\delta A \nabla_a \phi_0+\delta B\nabla_a\phi_1$. On the other hand{, if we set $ B $ to zero first and then vary the equation,} we find that $ \delta\ell_a=\delta A \nabla_a \phi_0 $. This variation is not valid, because variation of one degrees of freedom in metric has been killed in this relation{, i.e. $ \delta B\propto \ell^a\ell^b \delta g_{ab}=0 $.} This variation is responsible for  taking out the boundary from being null. {To avoid losing any degree of freedom, in the following}  we will not set $ B=0 $ until the end of calculations.   
 
 {We consider a  hypersurface  to be a leaf of one of the foliations,  let it   be $ \phi_0=const $.} 
 Variation of  HE action  on such  hypersurface  in the  double foliation determined by \cref{lfoliation,kfoliation,metric} leads to  \cite{Aghapour:2018icu}:
 \begin{align}
 \delta{ \mathcal S_{\scriptscriptstyle{HE}}}& =\tfrac{1}{8\pi}\,\delta\left(\int_{\mathcal N}\!\!\! \ud^{d-1} x \, \,\sqrt q [D(\Theta + \kappa)-B(\Xi+\bar{\kappa})] +  \int_{S_1}^{S_2}\!\!\! \ud^{d-2} x \,  \sqrt q \,\ln D\sqrt H\right)\nonumber \\
 &+\tfrac{1}{16\pi} \int_{\mathcal N}\!\!\! \ud^{d-1} x \,  \sqrt q\left[D(\Theta^{ab} -q^{ab}\,(\Theta + \kappa))-B(\Xi^{ab}-q^{ab}(\Xi+\bar{\kappa}))\right]\delta q_{ab}\nonumber\\  & + 2\,\omega^a\,\delta \beta_{1a}  -2 \Xi\, \delta B+2\Theta \delta D +\tfrac{1}{16\pi} \int_{S_1}^{S_2}\!\!\! \ud^{d-2} x\sqrt q (\ln D\sqrt H q^{ab})\delta q_{ab}  	\label{ThetaBnullfinal}, \,
 \end{align}
 where $ H $ is the determinant of $ H_{ij} $ defined above. The first line as a total variation suggest{s} appropriate boundary term that must be subtracted in order to have a well-posed variational principle. 
 The total variation term in the boundary  integral{, i.e. }
 \begin{equation}\label{GH}
 \sqrt{q} [D(\Theta + \kappa)-B(\Xi+\bar{\kappa})],
 \end{equation}
 has a nice geometric interpretation. It can be rewritten as $\sqrt{-g}\ v_a \mathscr K^a $, where $ v_a=\nabla_a\phi_0 =\frac{1}{\sqrt{H}}(D\ell_a-Bk_a)$ is the normal to the hypersurface, and  $ \mathscr K^a{}_{bc}=-(k^a\nabla_b\ell_c+l^a\nabla_bk_c) $ is defined  such that $ \mathscr K^a $ is its trace on two last indices\footnote{We note that the common Gibbons-Hawking-York term can be rewritten as: $ \sqrt{|h|}K=\sqrt{-g}\ \nabla_a\phi \  (n^a K) $, with $ K=-\nabla_a n^a $, in this sense, the above term is a generalization of GHY term  for null surfaces.}.
 
 On a null boundary now we set $ B=0 ${; thus,} we find the action with a well-posed variational principle to be
 \begin{equation}\label{null boundary term}
 { \mathcal S}=\tfrac{1}{16\pi}\int_{\mathcal{M}} \ud^d x \sqrt{-g}R-\tfrac{1}{8\pi}\int_{\mathcal{N}} \ud^{d-1} x \sqrt{q}[D(\Theta + \kappa)]+\tfrac{1}{8\pi}\int_{S_1}^{S_2}\ud^{d-2} x\sqrt{q}\ln D\sqrt{H}.
 \end{equation}
 
 Let us remind that{, as mentioned in previous subsection,} there is a scaling boost gauge symmetry in null hypersurface description by two vectors $ \ell_a $ and $ k_a $. We can use this gauge freedom to set $ D=1 $ { in this gauge and by putting $ B=0 $ we have $ \sqrt{H}=A $. Therefore,} the action becomes:
 \begin{equation}\label{actionNull}
 \mathcal{S}=\tfrac{1}{16\pi}\int_{\mathcal{M}} \ud^d x \sqrt{-g} \,R-\tfrac{1}{8\pi}\int_{\mathcal N}\!\!\! \ud^{d-1} x \, \,\sqrt q (\Theta + \kappa) +\tfrac{1}{8\pi}  \int_{S_1}^{S_2}\!\!\! \ud^{d-2} x \,  \sqrt q \,\ln A.
 \end{equation}
 This is what found  in  \cite{Lehner:2016vdi,Hopfmuller:2016scf}.
 In \cite{Parattu:2015gga} the authors also have set the lapse $ A=1 $, so the last term vanishes.
 %
 {Using the above action, the} variation of the principal function on the null boundary becomes:
 \begin{align}\label{principalNull}
 \delta\mathcal{S}=&\tfrac{1}{16\pi} \int_{\mathcal N}\!\!\! \ud^{d-1} x \,  \sqrt q\left(\left[(\Theta^{ab} -q^{ab}\,(\Theta + \kappa))\right]\delta q_{ab} + 2\,\omega_a\,\delta \beta_{1}^a  -2 \Xi\, \delta B\right)\nonumber\\	& +\tfrac{1}{16\pi} \int_{S_1}^{S_2}\!\!\! \ud^{d-2} x\sqrt q (\ln{A}\ q^{ab})\delta q_{ab}.  
 \end{align}  
 Similar expression for canonical structure on the null boundary has been found in \cite{Hopfmuller:2016scf}. There, the authors considered only the variations that keeps the boundary to remain null, {thus, the last term of the} first line was missing  in their analysis.   
 Note that in this expression $ \delta q_{ab} $, $ \delta\beta_{1a} $ and $ (-2\delta B)$ are all variations of the  metric components tangential to the boundary as easily can be seen from the decomposition \eqref{metric}. This is an important {point because, as in the non-null case, the variational principle tells us that for Dirichlet boundary condition} we  only need to fix the tangential metric components. Here also the number of  degrees of freedom match{es} with the non-null case{; for example,} in four dimensions $ \delta h_{ab} $ has six components {while}  $ (\delta q_{ab},\delta\beta_{1a},-2B) $ altogether have $ 3+2+1=6 $ components.
 \subsection{The stress tensor} 
 Having found the variation of action on the null boundary, we can use BY prescription to find {the} stress tensor on the boundary.
 In doing so we must differentiate the  principal function in space-time region, illustrated in Fig \ref{fig2}, with respect to the metric component tangential to the null boundary segments  $ \mathcal{N} $. Note that differentiating with respect to metric components in $ \Sigma_{1} $ and $ \Sigma_{2} $  gives the canonical momenta $ P^{ab} $ similar to the previous section.  According to the Eq.\eqref{metric}, the metric components tangential to $ \mathcal{N} $  are $ (\delta q_{ab},\delta\beta_{1a},-2B) $. Note that, in contrast to previous timelike boundary presented in the last section, there is no induced metric on null surface; {thus,} we must differentiate with respect to each component of metric separately according to:  
 \begin{equation}\label{DT}
 \frac{2}{\sqrt{q}}g_{ac}g_{bd}\frac{\delta{\mathcal{S}}|_{\mathcal N}}{\delta{g}_{cd}}=\frac{2}{\sqrt{q}}(q_{ac}q_{bd}\frac{\delta{\mathcal{S}}}{\delta{q}_{cd}}-q_{c(a}{}k_{b)}\frac{\delta{\mathcal{S}}}{\delta{\beta_{1c}}}+k_ak_b\frac{\delta{\mathcal{S}}}{\delta{(-2B)}} ).
 \end{equation}
 Where the notation $ |_{\mathcal N} $ means differentiating with respect to the metric component tangential to the null boundary. 
 On the other hand, the variational principle does not fix the boundary action completely{; therefore,} we  can subtract any functional $ {\mathcal S}_0 $  of fixed boundary data from the action \eqref{actionNull}. Using the expression \eqref{principalNull},{ hence, the} stress tensor components read: 
 \begin{align}
 \epsilon&\equiv\ell^{a}\ell^{b}T_{ab}=-\frac{1}{\sqrt{q}}\frac{\delta{\mathcal{S}}}{\delta{B}}=\tfrac{1}{8\pi}\big[\Xi-\Xi_0\big]\label{energydensity},\\
 j^a&\equiv q^{ac}\ell^{b}T_{cb}=\frac{1}{\sqrt{q}}\frac{\delta{\mathcal{S}}}{\delta{\beta_{1a}}}=\tfrac{1}{8\pi}\big[\omega^a-\omega^a_0\big]\label{momentomdesity},\\
 s^{ab}&\equiv q^{ac}q^{bd}T_{cd}=\frac{2}{\sqrt{q}}\frac{\delta{\mathcal{S}}}{\delta{q}_{ab}}=\tfrac{1}{8\pi}\big[\Theta^{ab} -q^{ab}\,(\Theta + \kappa)-\frac{2}{\sqrt{q}}\frac{\delta{\mathcal{S}_0}}{\delta{q}_{ab}}\big]\label{Sstres},
 \end{align} 
 where $ \omega^a_0 $ and $ \Xi_0 $ are defined as:
 \[  \omega^a_0=\frac{1}{\sqrt{q}}\frac{\delta{\mathcal{S}_0}}{\delta{\beta_{1a}}}\quad,\quad \Xi_0=-\frac{1}{\sqrt{q}}\frac{\delta{\mathcal{S}_0}}{\delta{B}}\]
 According to BY prescription,  \cref{energydensity,momentomdesity,Sstres},  may be used {to define} quasilocal energy density $ \epsilon $, tangential momentum density $ j^a $ and spatial stress $ s^{ab} $, respectively.
 
 The subtraction term $ \mathcal{S}_0 $ is related to {the zero point of energy and  must be well chosen so} that the stress tensor leads to finite energy for different systems and conventionally zero energy for Minkowski space time.  The  expression \eqref{energydensity} is similar to the Eq. \eqref{QLE} for timelike boundary. $ \Xi$ is defined by $ \Xi=-q^{ab}\nabla_ak_b $ while $ \mathbf{k} $ {reads as} $ \mathbf{k}=-q^{ab}\nabla_an_b $. The similarity is in the sense that for null boundary the only  vector pointing out of the boundary is $ k_a $. For non-null boundaries{,} studied in the previous section, the  normal to boundary was also pointing out of it{; however, note that for null hypersurfaces} the normal is within the boundary. 
 \section{Quasi-local quantities }\label{charge}
 In this section we examine the proposed stress tensor by evaluating conserved charges of various {space-times and then we} compare it with the known results.
 Here, as to the timelike boundary, we propose the Quasi-local quantity to be:
 \begin{equation}\label{Qchargenull}
 Q_{\xi}=\int_{S}d^{d-2}x\sqrt{q}\ T_{ab}\,\ell^a\xi^b\ .
 \end{equation}
 { where $ \ell^a $ has the role of time flow on the boundary while  $ \xi^a $ is  a Killing vector which generates} an isometry of the boundary. If $ \ell^a $ {is} the generator of time translation symmetry, then the total energy becomes: 
 \begin{equation}\label{Energy}
 E=\int_{S}d^{d-2}x\sqrt{q} \ \epsilon ,
 \end{equation}
 where the quasi-local  energy density $ \epsilon $ is defined by \eqref{energydensity}. When there is a rotational Killing vector $ \zeta^a $, then its corresponding angular momentum is:
 \begin{equation}\label{angularmomentum}
 J=\int_{S}d^{d-2}x\sqrt{q} \ j_a \zeta^a.
 \end{equation} 
 In the following we will calculate the above  quantities for some well-known gravitational solutions. 
 \subsection{Minkowski space} 
 The simplest  example for investigation  is Minkowski spacetime. In retarded-spherical coordinates the metric is as follows:  
 \begin{equation}\label{Minkowski}
 ds^2=-du^2-2dudr+r^2d\Omega^2
 \end{equation}
 where  $ d\Omega^2=\gamma_{AB}d\sigma^Ad\sigma^B $ is the metric on a unit sphere. 
 It is evident that $ u=const. $ is a null surface. Comparing the above line element with Eq.\eqref{metric} and using the foliation relations \eqref{lfoliation} and\eqref{kfoliation} yields $ \ell_a=\nabla_a u $, $ k_a=\nabla_ar+\tfrac12\nabla_au $ and $ q_{AB}=r^2\gamma_{AB} $. From the definitions \eqref{BGOdef} one easily finds:
 \begin{equation}
 \Theta_{AB}=-2\Xi_{AB}=r\gamma_{AB},\quad\omega_A=0.
 \end{equation}
 As a result $ \Xi=-\frac{1}{r} $, and the integral $ \int d\theta d\phi \sqrt{q}\ \Xi=-\int d\theta d\phi\ r\sin^2\theta $  will be infinite as $ r\to\infty $. Subtracting $ \Xi_0 $  leads to zero energy for flat space as expected. { In fact, by using its definition we calculate $ \Xi_0$ as embedding of boundary in flat space. Then, one could show that $ \Xi=\Xi_0 $ and as a result  the energy density vanishes. In other words, in this case, the physical and reference spacetimes are the same.  }
 
 \subsection{Schwarzschild black hole}
 Our next simple example is Schwarzschild black hole. In retarded  Eddington{-}Finkelstein coordinates the metric is given by:
 \begin{equation}\label{Schwarzschild}
 ds^2=-f(r)du^2-2dudr+r^2d\Omega^2
 \end{equation}
 In this case we have $ \ell_a=\nabla_a u $ and $ k_a=\tfrac{1}{2f(r)}\nabla_au+\nabla_ar $. 
 Here, by calculating the quasilocal energy density, we get:
 \begin{equation}\label{energySchw}
 \epsilon=\frac1{8\pi}[\Xi-\Xi_0]= \frac1{8\pi}[-\frac{f(r)}{r}+\frac{1}{r}].
 \end{equation}
 { Here, we have used the fact $ \Xi_0=-\frac{1}{r} $, as a result of embedding in flat space, and  $ \Xi=-\frac{f(r)}{r} $ which can easily be found.} By replacing $ f(r)=1-\frac{2M}{r} $, we obtain $ \epsilon=\frac{ M}{4\pi r^2} $. Thus the total energy becomes:
 \begin{equation}\label{Mass}
 E=\int_0^\pi \int_0^{2\pi} d\theta d\phi\ r^2 \sin\theta\ \epsilon = 4\pi  r^2 (\frac{ M}{4\pi r^2}) =M.
 \end{equation}
 It is interesting that the quasilocal energy, as calculated for the null observers in $ \mathcal{N} $, is independent of the  distance $ r $ for which the  energy is calculated. 
 At first sight, it may seems  strange  because the usual BY expression for energy is just equal to the ADM mass at $ r\to\infty $. {However, notice that as mentioned previously,} in general the function $ E(r) $ is observer dependent.  The in-dependency of energy to the distance also  has been observed  for  {boosted} foliation of  Schwarzschild black hole in \cite{Brown:2000dz}.
  \subsection{AdS-Schwarzschild black hole}
   In the above example{ ,} we considered { the} usual asymptotic flat Schwarzschild black { hole}. { The validity of the above procedure for asymptotic AdS/dS black holes could be justified as follows.  In fact, our method is similar to the usual Brown-York strategy} for quasi-local quantities. Let us point { out} that in timelike case{ , for}  every geometry that can be contained in a spacetime region depicted in Fig.\ref{fig1}, one must be able to calculate  quasi-local quantities. { The only subtlety here is the problem of}  choosing the reference spacetime. For  asymptotic flat space{ ,} the natural choice is Minkowski space{ . On the other hand,} the preferred   reference for  asymptotic AdS spacetime is the AdS space (see e.g \cite{Brown:1994gs}). For  spacetime region in this study, namely Fig.\ref{fig2}, the same story is true. { Moreover,}  for AdS-Schwarzschild black hole we have to embed the null hypersurface in AdS space in order to obtain { an} expression for { the } reference term. Consider the metric 
   \begin{equation}\label{adsSchwarzschild}
  ds^2=-f(r)du^2-2dudr+r^2d\Omega^2
  \end{equation}
   with $ f(r)=1+\tfrac{r^2}{l^2}-\tfrac{2M}{r} $, for AdS-Schwarzschild black hole. One can easily calculate $ \Xi_0 $ and $ \Xi $ as
   \begin{equation}\label{adsxi}
   \Xi_0=-\tfrac{1}{r}-\tfrac{r}{l^2},\quad\Xi=-\tfrac{f(r)}{r}
   \end{equation}
{  from which the} quasilocal energy density will be obtained as $ \epsilon=\frac{ M}{4\pi r^2} $.  Thus, simple integration leads to the ADM mass $ M $ for these black holes.  Furthermore, the same results could be derived in the case of asymptotic dS black hole  if we choose the reference spacetime to be dS. 
 
 \subsection{Slow-rotating  black hole}
 Here we examine conserved quantities for slow rotating Kerr black holes. Again we write the metric in the retarded  Eddington{-}Finkelstein coordinates:
 \begin{equation}\label{Kerrmetric}
 ds^2=-f(r)du^2-2dudr+r^2d\Omega^2+ \frac{2J}{r}\sin^2\theta \ du d\phi+\frac{2J}{rf(r)}\sin^2\theta \ dr d\phi
 \end{equation} 
 Comparing with metric \eqref{metric} we arrive at: 
 \begin{equation}\label{comps}
 A=1,\quad\ B=0,\quad C=\frac{1}{2f},\quad D=1,\quad \beta_{1\phi}=\frac{2J}{r}\sin^2\theta,\quad\beta_{2\phi}=\frac{2J}{r{ }f}\sin^2\theta.
 \end{equation}
 Up to the first power in  $J$, one finds the same expression for energy as the Schwarzschild case. The angular momentum quantity is related to $ \omega_a $, using its definition we find:
 \begin{equation}\label{omega}
 \xi^a\omega_a=\frac{3J\sin^2\theta}{r^2},
 \end{equation}
 where $  \xi=\pd_\phi $ is rotational Killing symmetry.
 Therefore, the total angular momentum becomes:
 \begin{equation}\label{J}
 Q_{\xi}=\frac{1}{8\pi}\int_0^\pi \int_0^{2\pi} d\theta d\phi\ r^2 \sin\theta\frac{3J\sin^2\theta}{r^2}=J
 \end{equation}
 
 \subsection{Asymptotic flat spacetime and the Bondi mass}
 Here, we want to study the quasi-local quantities for gravitational theories in which the metric  { has an asymptotic flat space behavior}. In order to do that, we suppose the metric in the Bondi coordinates{. In} this gauge, the most general four-dimensional metric takes the
 form:
 \begin{equation}\label{bondimetric}
 ds^2=-U V du^2-2V dudr+ q_{AB}(d\sigma^A + U^A\,du)(d\sigma^B + U^B\,du)
 \end{equation}
 where $ \pd_r\det(\frac{ q_{AB}}{r^2})=0 $. By comparison the above equation  with Eq.\eqref{metric}, we find:
 \begin{equation}\label{bondicoef}
 A=V,\quad B=0,\quad C=\frac{U}{2},\quad D=1,\quad \beta_{0}^A=U^A,\quad \beta_{1}^A=0.
 \end{equation}
 The expressions for $ \Theta_{AB} $ and $ \Xi_{AB} $ in the metric \eqref{metric} have been calculated explicitly in \cite{Aghapour:2018icu} and are:
 \begin{align}
 \Theta_{AB} &=- \frac1{2\sqrt H}\,(B\,\partial_0q_{AB} -A\,\partial_1q_{AB} -2\,B\,\mathcal D_{(A} \beta_{0 B)}+2\,A\,\mathcal D_{(A} \beta_{1 B)})\label{theta}\\
 \Xi_{AB} &= -\frac1{2\sqrt H}\,(-D\,\partial_0q_{AB} +C\,\partial_1q_{AB} +2\,D\,\mathcal D_{(A} \beta_{0 B)}-2\,C\,\mathcal D_{(A} \beta_{1 B)})\label{Xi}
 \end{align}
 {where} $ \mathcal D $ is covariant derivative on two sphere $ \mathcal{S} $, compatible with the metric $ q_{AB} $ \footnote{These relations are counterpart to the well-known relation for extrinsic curvature in $3+1$ decomposition: $ K_{ij}=-\frac{1}{2N}(\pd_th_{ij}-2  D_{(i} N_{j)} ) $}. Using standard asymptotic expansions \cite{Bondi:1962px}:
 \begin{align*}
 U=1-\frac{2m_B}{r}+{\mathcal O}(\frac{1}{r^2}),\quad V=1+{\mathcal O}(\frac{1}{r^2}) \nonumber\\
 \beta_{0}^A=\frac{W^A}{r^2}+{\mathcal O}(\frac{1}{r^3}),\quad q_{AB}=r^2\gamma_{AB}+{\mathcal O}(r)
 \end{align*}
 we can easily find the following leading terms for $ \Xi $:
 \begin{equation*}
 \Xi=-\frac{1}{r}+\frac{2m_{B}(u,\sigma^A)}{r^2}+\frac{{\mathcal D}_{A}W^A}{r^2}+{\mathcal O}(\frac{1}{r^3})
 \end{equation*}
 By embedding in flat space we have the reference term $ \Xi_0=-\frac{1}{r} $. The term $ {\mathcal D}_{A}W^A $ is a total derivative on compact two sphere $ \mathcal{S} $ {which vanishes} by integration. {Thus,} finally the total energy becomes:
 \begin{equation}
 E=\frac{1}{8\pi}\int_{S}d^{2}x\sqrt{q} \ \epsilon =\frac{1}{4\pi}\int_{S}d\Omega\  m_{B}(u,\sigma^A)
 \end{equation}
 {which} is the expression known as Bondi mass.
 \subsection{A possible counterterm for asymptotic flat spacetime }
 We have seen that for  asymptotic AdS space there is counterterm method for regularizing quasi-local quantities that has some advantages. The {first benefit is that the} dependency on the reference spacetime and  mathematical difficulty of embedding is avoided. Furthermore{,} the counterterms have direct interpretation in the dual field theory  and has important role for building  the dictionary of AdS/CFT. {Then the} natural question is that: why  there is {no}  similar counterterms  for flat space?
 For flat space there is not a length scale counterpart to the AdS length scale{; thus,} we can not have terms like: $ \ell^{2n-1}\mathcal{R}^n$. { In addition,} 
 the counterterms should not spoil the variational principle { so that the} extrinsic curvature terms are forbidden. Therefore{,} on the time-like boundary we don't have a { viable candidate as}  counterterm in asymptotic flat spacetime. {However, as we have seen in this paper,} null boundaries are special.
 Consider the following term on the null boundary:
 \begin{equation}
 \alpha\int_{\mathcal{N}} \ud^{d-1} x \sqrt{q}\  B \Theta
 \end{equation}
 with  arbitrary numerical coefficient $ \alpha $. The first point is that adding such term is compatible with Dirichlet boundary conditions. Also{,  we must note that although the above term vanishes on the null boundary, } its variation is not zero. {Therefore, it has a contribution to the stress tensor and energy  though  it is zero for on-shell action.} 
 Fortunately{,} we can set  
 the coefficient  $ \alpha=-\frac12 $ so that its contribution to energy make the total finite. 
 By adding this term to the boundary action, the quasilocal energy density becomes:
 \begin{equation}
 \epsilon=\tfrac{1}{8\pi}\big[\Xi+\frac{1}{2}\Theta]\label{energyfinite}
 \end{equation}
 For asymptotic flat spacetime studied previously, using \eqref{theta}, $ \Theta $ {is} easily  computed and its leading terms {is} as:
 \begin{equation}\label{asymtheta}
 \Theta=\frac{2}{r}+\mathcal{O}(\frac{1}{r^3}).
 \end{equation}
 Therefore the expression \eqref{energyfinite} leads to  correct total energy without needing to embedding and reference spacetime. 
 \section{Conclusion and outlook }
 In this {work} we have extended the Brown-York prescription to the case of null boundaries.  {For the achievement of this goal we needed a general double foliation  framework. The mathematics of this general double foliation  is described in \cite{Aghapour:2018icu} and reviewed and clarified here. The main reason for considering   such framework is that a single or double null foliation is basically gauge fixed, and are not appropriate for a variational  problem,  because some degrees of freedom are already fixed by considering a null foliation:
  \[ g^{\phi\phi}=g^{-1}(d\phi,d\phi)=0. \]Variation of Hilbert-Einstein  is calculated on such general double foliation in \cite{Aghapour:2018icu}. The main unanswered question in that article was to determine the physical meaning of metric variation on such null boundary. This question is answered here.} Especially we have seen that {the} variation  of one metric  component  is responsible for taking away the boundary from being null. {It was shown that an exact derivation} with this variation gives the expression for quasilocal energy density. {In addition, the  expressions for total energy and angular momentum were} examined for some known spacetimes.   {Moreover, a special property of the calculated energy for null observers was found to be the { in-dependency}} to radial distance. {Furthermore, it  was shown  that in the case of null boundaries, there is a possible counterterm which is consistent with variational principle and can be added to boundary action  so that} the total energy becomes finite, without the necessity of embedding in reference spacetime. 
 
 Regarding the similarity with the  standard AdS/CFT dictionary, one may interpret the   introduced  stress tensor as expectation value of stress tensor in dual field theory to flat space. {Because, according to Penrose diagrams for  asymptotic flat spacetimes, the null hypersurfaces $ {\mathscr I}^{+} $ and  $ {\mathscr I}^{-} $ are regarded as the boundaries of spacetime; thus,} one application of the stress tensor found in this paper may be flat holography. { In this context, the relation between  asymptotic symmetries and some  special limits - for example, Carrolian symmetry -} is worth to investigate\cite{Ciambelli:2018ojf}. 
 
 {Another} application of the formalism presented in this article and \cite{Aghapour:2018icu} {is} to revisit the gravity in light-front. The usual investigation of gravity in light-front {uses the} double-null foliation of spacetime \cite{Alexandrov:2014rta,Shyam:2015vxa}. {However, as noted above, the} double-null foliation leads to partial gauge fixing of the metric. It is in contrast to field theory formulation  in the  light-front coordinate which no gauge fixing is required. As we have seen a general double foliation can preserve all degrees of freedom. {Thus,} it is motivating to revisit gravity in light-front.
 %

\subsubsection*{\bf Acknowledgments}
 The author would like to thank the theory group of School of Particles and Accelerators (TSPA), specially { A. Naseh} for a lot of valuable discussions and insightful comments. The author is also very grateful to { S. Aghapour} for a lot of discussions during previous collaboration{,  H. Shenavar and  F. Hopfmüller} for  reading the manuscript and  helpful comments.  I
 also thank the anonymous referee, for suggestions that leads to clarification of some ideas.
 

 \bibliographystyle{hunsrt}
 \bibliography{NullStressTensor}
\end{document}